\documentclass[
reprint,
superscriptaddress,
nofootinbib,
nobibnotes,
amsmath,amssymb,
aps,
prl,
floatfix
]{revtex4-2}
\usepackage{graphicx}
\usepackage{dcolumn}
\usepackage{bm}
\usepackage{amssymb}
\usepackage{hyperref}
\usepackage{appendix}
\usepackage{float}
\begin{document}

\title{Experimental Evidence of Vortex $\gamma$ Photons in All-Optical Inverse Compton Scattering}

\author{Mingxuan Wei}
\thanks{These authors have contributed equally to this work.}
\author{Siyu Chen}
\thanks{These authors have contributed equally to this work.}
\affiliation{State Key Laboratory of Dark Matter Physics, Key Laboratory for Laser Plasmas (MoE), School of Physics and Astronomy, Shanghai Jiao Tong University, Shanghai 200240, China}

\author{Yu Wang}
\thanks{These authors have contributed equally to this work.}
\affiliation{Ministry of Education Key Laboratory for Nonequilibrium Synthesis and Modulation of Condensed Matter, State key laboratory of electrical insulation and power equipment, Shaanxi Province Key Laboratory
of Quantum Information and Quantum Optoelectronic Devices, School of Physics, Xi’an Jiaotong University, Xi’an 710049, China}

\author{Pei-Lun He}
\affiliation{State Key Laboratory of Dark Matter Physics, Key Laboratory for Laser Plasmas (MoE), School of Physics and Astronomy, Shanghai Jiao Tong University, Shanghai 200240, China}
\affiliation{Collaborative Innovation Center of IFSA, Shanghai Jiao Tong University, Shanghai 200240, China
}

\author{Xichen Hu} \author{Mingyang Zhu} \author{Hao Xu} 
\author{Weijun Zhou} \author{ Jiao Jia}
\affiliation{State Key Laboratory of Dark Matter Physics, Key Laboratory for Laser Plasmas (MoE), School of Physics and Astronomy, Shanghai Jiao Tong University, Shanghai 200240, China}

\author{Xulei Ge}
\affiliation{State Key Laboratory of Dark Matter Physics, Key Laboratory for Laser Plasmas (MoE), School of Physics and Astronomy, Shanghai Jiao Tong University, Shanghai 200240, China}
\affiliation{Tsung-Dao Lee Institute, Shanghai Jiao Tong University, Shanghai 201210, China}

\author{Lin Lu} 
\author{Boyuan Li} \author{Feng Liu}
\author{Min Chen}
\author{Liming Chen}
\affiliation{State Key Laboratory of Dark Matter Physics, Key Laboratory for Laser Plasmas (MoE), School of Physics and Astronomy, Shanghai Jiao Tong University, Shanghai 200240, China}
\affiliation{Collaborative Innovation Center of IFSA, Shanghai Jiao Tong University, Shanghai 200240, China
}
\author{Pavel Polynkin}
\affiliation{College of Optical Sciences, The University of Arizona, Tucson, Arizona 85721, USA}

\author{Jian-Xing Li}
\email{jianxing@xjtu.edu.cn}
\affiliation{Ministry of Education Key Laboratory for Nonequilibrium Synthesis and Modulation of Condensed Matter,  State key laboratory of electrical insulation and power equipment,  Shaanxi Province Key Laboratory
of Quantum Information and Quantum Optoelectronic Devices,  School of Physics, Xi’an Jiaotong University, Xi’an 710049, China}
\affiliation{Department of Nuclear Physics, China Institute of Atomic Energy, P.O. Box 275(7), Beijing 102413, China}

\author{Wenchao Yan}
\email{wenchaoyan@sjtu.edu.cn}
\affiliation{State Key Laboratory of Dark Matter Physics, Key Laboratory for Laser Plasmas (MoE), School of Physics and Astronomy, Shanghai Jiao Tong University, Shanghai 200240, China}
\affiliation{Collaborative Innovation Center of IFSA, Shanghai Jiao Tong University, Shanghai 200240, China
}

\author{Jie Zhang}
\affiliation{State Key Laboratory of Dark Matter Physics, Key Laboratory for Laser Plasmas (MoE), School of Physics and Astronomy, Shanghai Jiao Tong University, Shanghai 200240, China}
\affiliation{Collaborative Innovation Center of IFSA, Shanghai Jiao Tong University, Shanghai 200240, China
}
\affiliation{Tsung-Dao Lee Institute, Shanghai Jiao Tong University, Shanghai 201210, China}

\date{\today}
\begin{abstract}
Vortex $\gamma$ photons carrying orbital angular momenta (OAM)  hold great potential for various applications. However, their generation remains a great challenge. Here, we successfully generate sub-MeV vortex \(\gamma\) photons via all-optical inverse Compton scattering of relativistic electrons colliding with a sub-relativistic Laguerre-Gaussian laser. In principle, directly measuring the OAM of $\gamma$ photons is challenging due to their incoherence and extremely short wavelength. Therein, we put forward a novel method to determine the OAM properties by revealing the quantum opening angle of vortex $\gamma$ photons, since vortex particles exhibit not only a spiral phase but also transverse momentum according to the quantum electrodynamics theory. Thus, $\gamma$ photons carrying OAM manifest a much larger angular distribution than those without OAM, which has been clearly observed in our experiments. This angular expansion is considered as an overall effect lying beyond classical theory.
Our method provides the first experimental evidence for detecting vortex $\gamma$ photons and opens a new perspective for investigating OAM-induced quantum phenomena in broad fields.
\end{abstract}

\maketitle

Vortex photons carry orbital angular momenta (OAM) along their propagation direction \cite{yaoOrbitalAngularMomentum2011,allenOrbitalAngularMomentum1992}. The introduction of vortex laser adds an additional degree of freedom, leading to their widespread application in various fields, including particle manipulation \cite{heDirectObservationTransfer1995b,patersonControlledRotationOptically2001,rosales-guzmanReviewComplexVector2018,shenOpticalVortices302019}, quantum optics \cite{ficklerQuantumEntanglementHigh2012,mairEntanglementOrbitalAngular2001}, imaging \cite{swartzlanderAstronomicalDemonstrationOptical2008,swartzlanderPeeringDarknessVortex2001}, high-harmonic generation \cite{gauthierTunableOrbitalAngular2017}, etc. Recent advancements suggest that vortex $\gamma$ photons could probe the isovector giant quadrupole resonance without interference from dipole transitions, owing to the conservation of angular momentum, thus generating considerable interest in the field of nuclear physics \cite{luManipulationGiantMultipole2023a}. Vortex photons from visible to extreme ultraviolet ranges are routinely generated in experiments via spatial light modulators \cite{chattrapibanGenerationNondiffractingBessel2003}, spiral phase plates \cite{suedaLaguerreGaussianBeamGenerated2004a,peeleObservationXrayVortex2002}, or holograms \cite{heckenbergGenerationOpticalPhase1992}.
For the soft X-ray range, several generation methods have been proposed, including diffractive optics \cite{leeLaguerreGaussHermite2019}, helical undulators \cite{katohHelicalPhaseStructure2017,bahrdtFirstObservationPhotons2013}, and seed laser beams with OAM \cite{hemsingCoherentOpticalVortices2013,hemsingGeneratingOpticalOrbital2011a,hemsingEchoEnabledXRayVortex2012}. Additionally, studies on the helical motion of charged particles in lasers or undulators have demonstrated that single photons in harmonic radiation naturally carry OAM \cite{sasakiProposalGeneratingBrilliant2008a,katohAngularMomentumTwisted2017a,tairaGammarayVorticesNonlinear2017}. Due to the difficulty of manipulating angular momentum at high energies and preserving helical phase structure, the generation of high energy vortex $\gamma$ photons still remains a challenge.

The development of ultra-intense laser facilities \cite{dansonPetawattExawattClass2019,weberP3InstallationHighenergy2017a} has paved the way for enhancing photon energy into the hard X-ray or $\gamma$ rays range via inverse Compton scattering schemes \cite{sarriUltrahighBrillianceMultiMeV2014,yanHighorderMultiphotonThomson2017,khrennikovTunableAllOpticalQuasimonochromatic2015,mirzaieAllopticalNonlinearCompton2024,powersQuasimonoenergeticTunableXrays2014,chenMeVEnergyRaysInverse2013,milburnElectronScatteringIntense1963,dipiazzaExtremelyHighintensityLaser2012a,nakamuraHighPower$ensuremathgamma$RayFlash2012}. High-harmonic photons carry OAM when a circularly polarized (CP) laser interacts with relativistic electrons. This can be interpreted as the helical motion of an electron within a CP laser field under a classical framework \cite{katohHelicalPhaseStructure2017,sasakiProposalGeneratingBrilliant2008a}.  The generation of vortex $\gamma$ photons via spin-to-orbital angular momentum conversion in strong-field quantum electrodynamics processes was theoretically established through Furry scattering theory \cite{ababekriVortexPhotonGeneration2024}, manifesting as quantized phase singularities in radiation patterns.
However, the rotational symmetry of CP lasers reduces the yield as the harmonic order increases, limiting the efficiency of generating $\gamma$ photons \cite{liSpintoorbitalAngularMomentum2020a}.

To surmount the aforementioned difficulties, an inverse Compton scattering process utilizing Laguerre–Gaussian (LG) lasers, which carry OAM, has been proposed \cite{petrilloComptonScatteredXGamma2016}. This approach, owing to angular momentum conservation, enables generating a large number of $\gamma$ photons with OAM inherited from the LG laser. The theoretical feasibility of generating twisted radiation through inverse Compton scattering \cite{zhouGammarayVortexBurst2023,chenGenerationTwistedRay2019,jentschuraComptonUpconversionTwisted2011,jentschuraGenerationHighEnergyPhotons2011b,stockComptonScatteringTwisted2015,liuTwistedRadiationNonlinear2020}, as well as schemes for producing $\gamma$ rays carrying beam OAM as a collective effect of individual photons, have been proposed numerically \cite{liuGenerationGammarayBeam2016,chenGRayBeamsLarge2018,fengEmissionGRayBeams2019,zhangEfficientBrightGray2021,huAttosecondGrayVortex2021}. Notably, due to the incoherent and  extremely short wavelength of $\gamma$ rays, optical measurement methods based on coherence become inapplicable. Therefore, the detection of high-energy $\gamma$ photons with OAM in experiments also remains a great challenge.

\begin{figure}
    \centering
    \includegraphics[width=\linewidth]{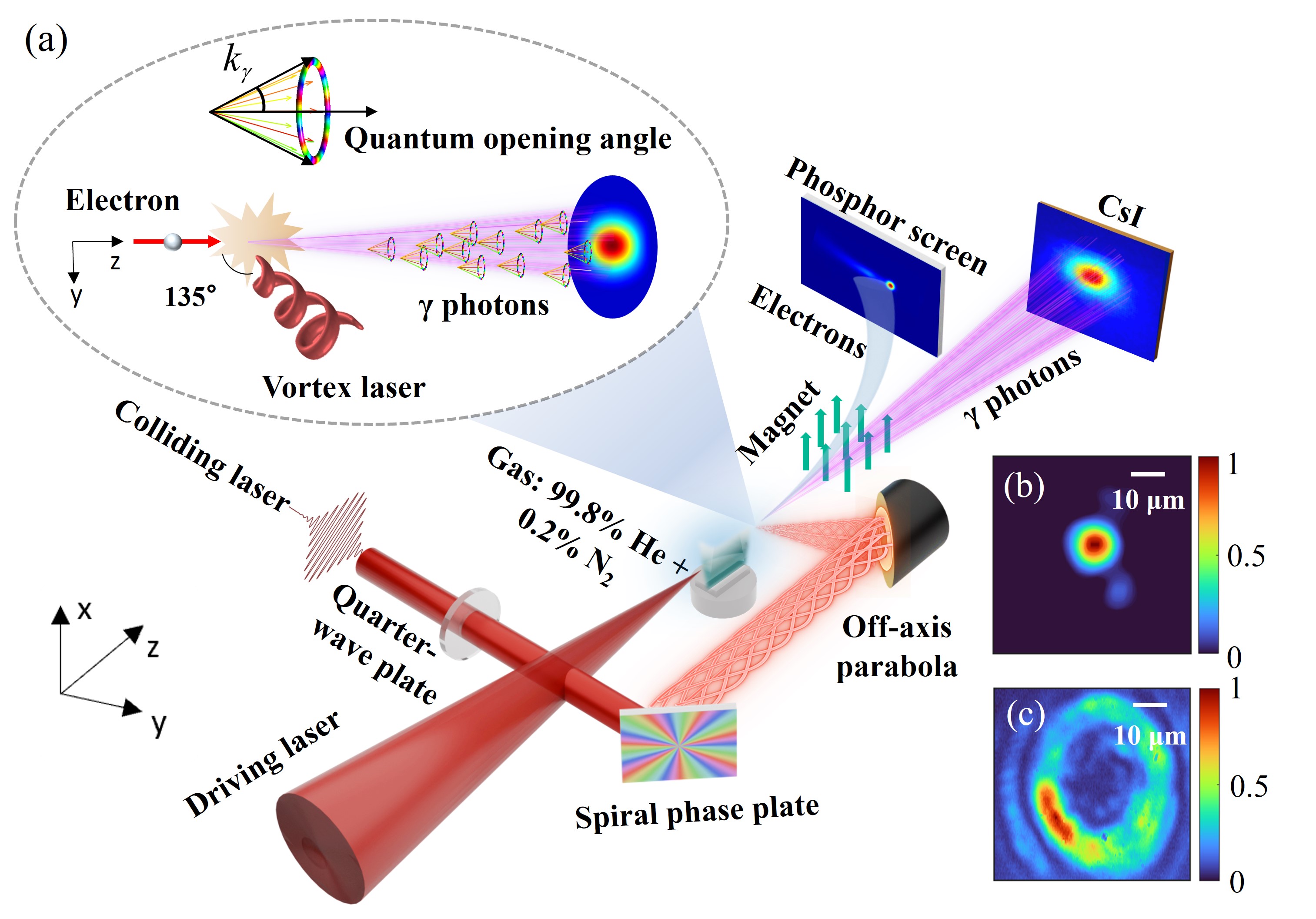}
    \caption{(a) Experimental layout. The electron beam, produced via the LWFA, propagated in the $+\hat{z}$ direction. The phosphor screen is used to measure energy spectrum of electron, while the angular distribution of $\gamma$ rays is measured by CsI. The spiral phase plate, positioned before the OAP, was used to generate vortex laser. Both the driving laser and colliding laser were horizontally polarized in the $\hat{y}-\hat{z}$ plane. The polarization state of the colliding laser is controlled using a quarter-wave plate. The upper-left inset illustrates the interaction principle between relativistic electrons and vortex laser at a collision angle of 135°. $k_\gamma$ is the momentum vector of the vortex photon. (b) and (c) are the focal spot patterns of the Gaussian and LG colliding laser, respectively.}
    \label{fig1}
\end{figure}

\begin{figure}
    \centering
    \includegraphics[width=\linewidth]{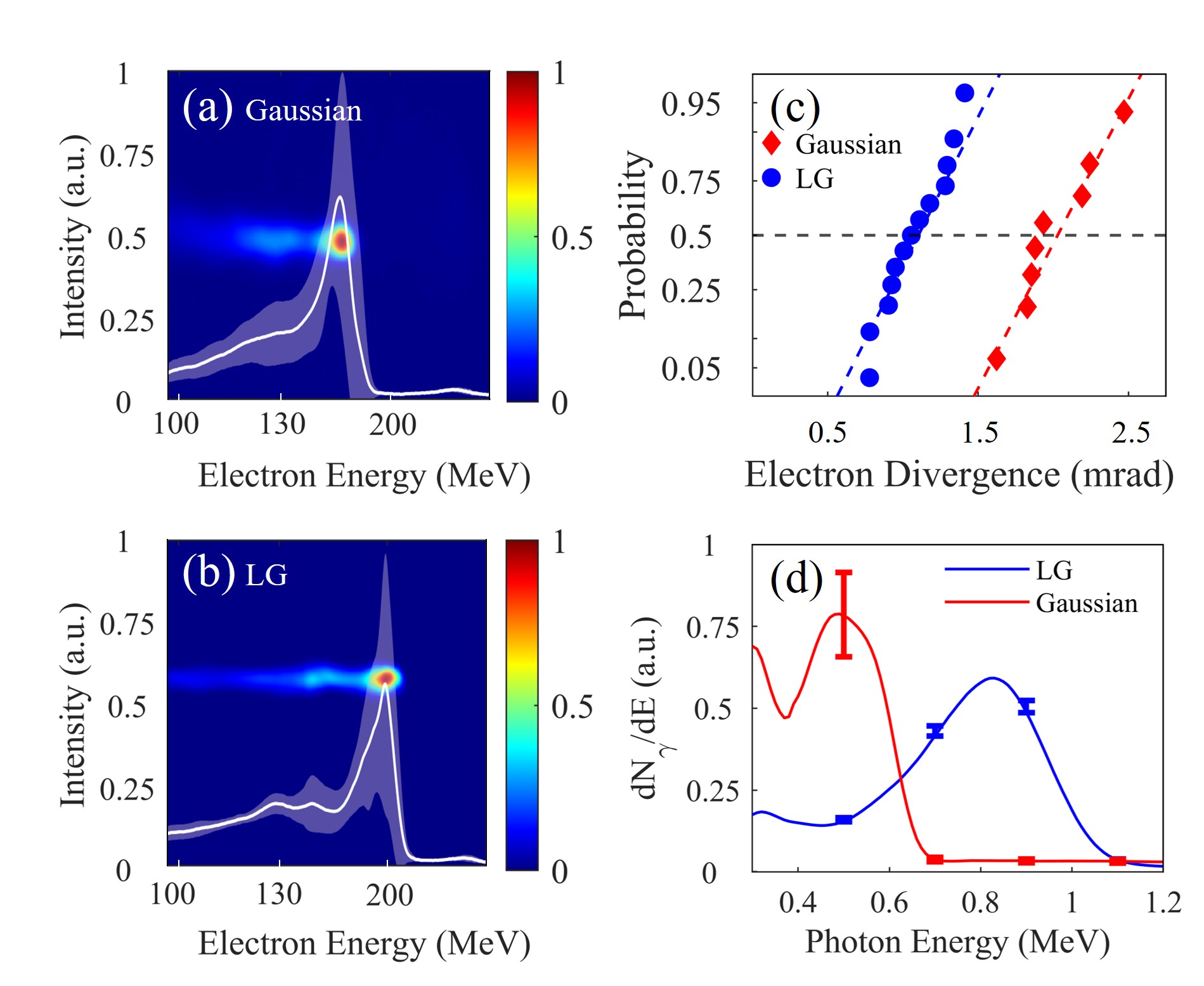}
    \caption{Experimental Data. The typical energy spectra of electron beams interacting with (a) Gaussian laser, and (b) LG laser, respectively. The shaded areas in (a) and (b) represent the error bands of the electron energy spectra of all shots under the corresponding colliding lasers, and the white lines show the average spectra in two cases. (c) Probability plot of statistical transverse divergence angle distributions of the electron beams (blue circle: LG; red diamond: Gaussian). The blue and red dashed lines correspond to the normal distribution fitting of the electron divergence angles in two cases, whereas the black dashed line represents the average value. (d) Corresponding $\gamma$ rays spectra under different colliding laser configurations. Other experimental parameters are the same as those in Fig. \ref{fig1}.}
    \label{fig2}
\end{figure}

\begin{figure}
    \centering
    \includegraphics[width=0.9\linewidth]{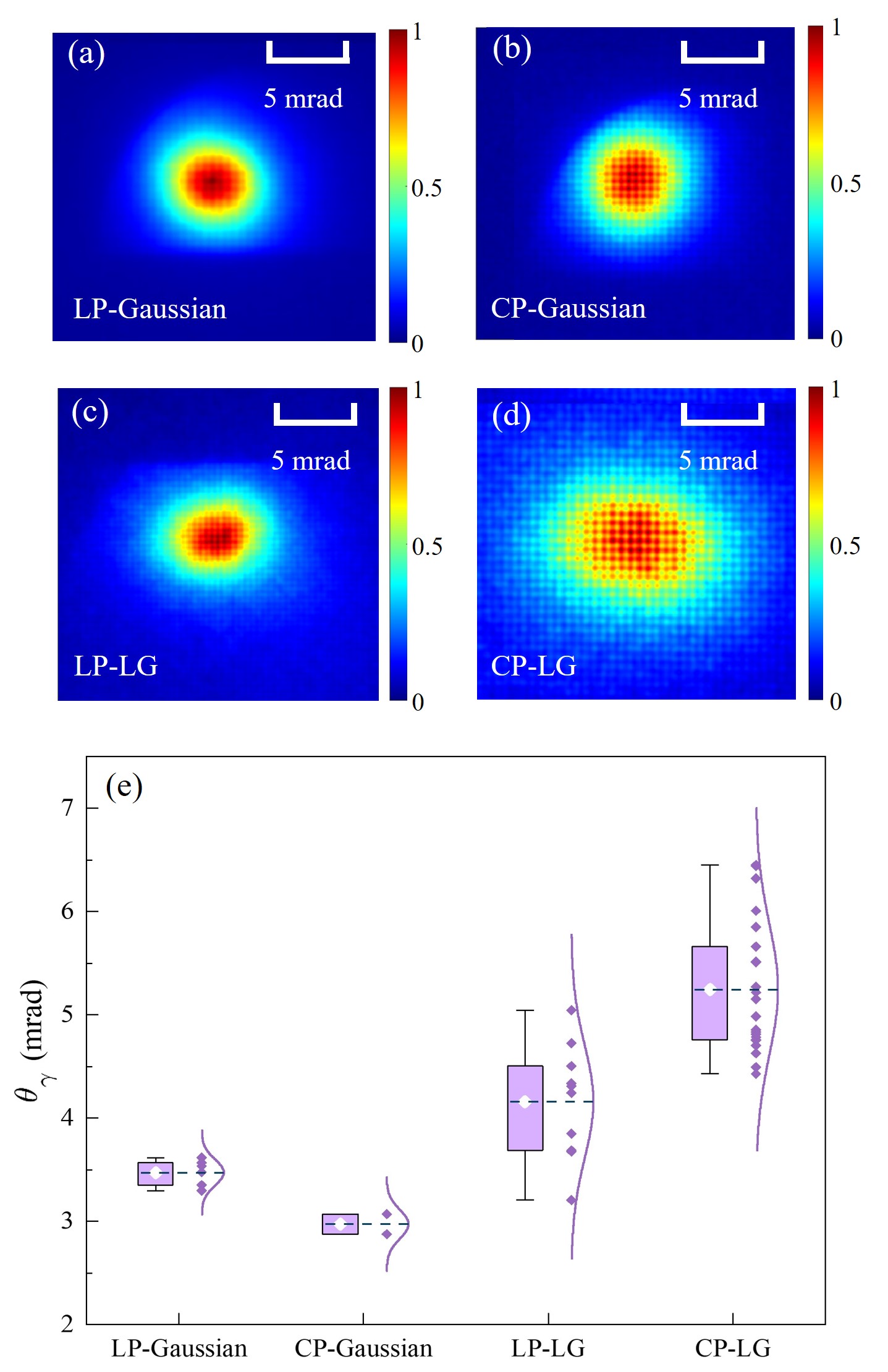}
    \caption{Angular distribution of $\gamma$ rays generated by four colliding laser configurations: (a) LP-Gaussian, (b) CP-Gaussian, (c) LP-LG, and (d) CP-LG.  Panel (e) quantifies the statistical distribution of divergence angles through box-whisker analysis. The purple dots show one shot data, and the purple curve represents the fitting of all data using a normal distribution. White diamonds denote mean values, with box boundaries marking the 25th and 75th percentiles.  Whiskers extend to 1.5 times the interquartile range. Other experimental parameters remain consistent with Fig. \ref{fig1}.}
    \label{fig3}
\end{figure}

\begin{figure}
    \centering
    \includegraphics[width=\linewidth]{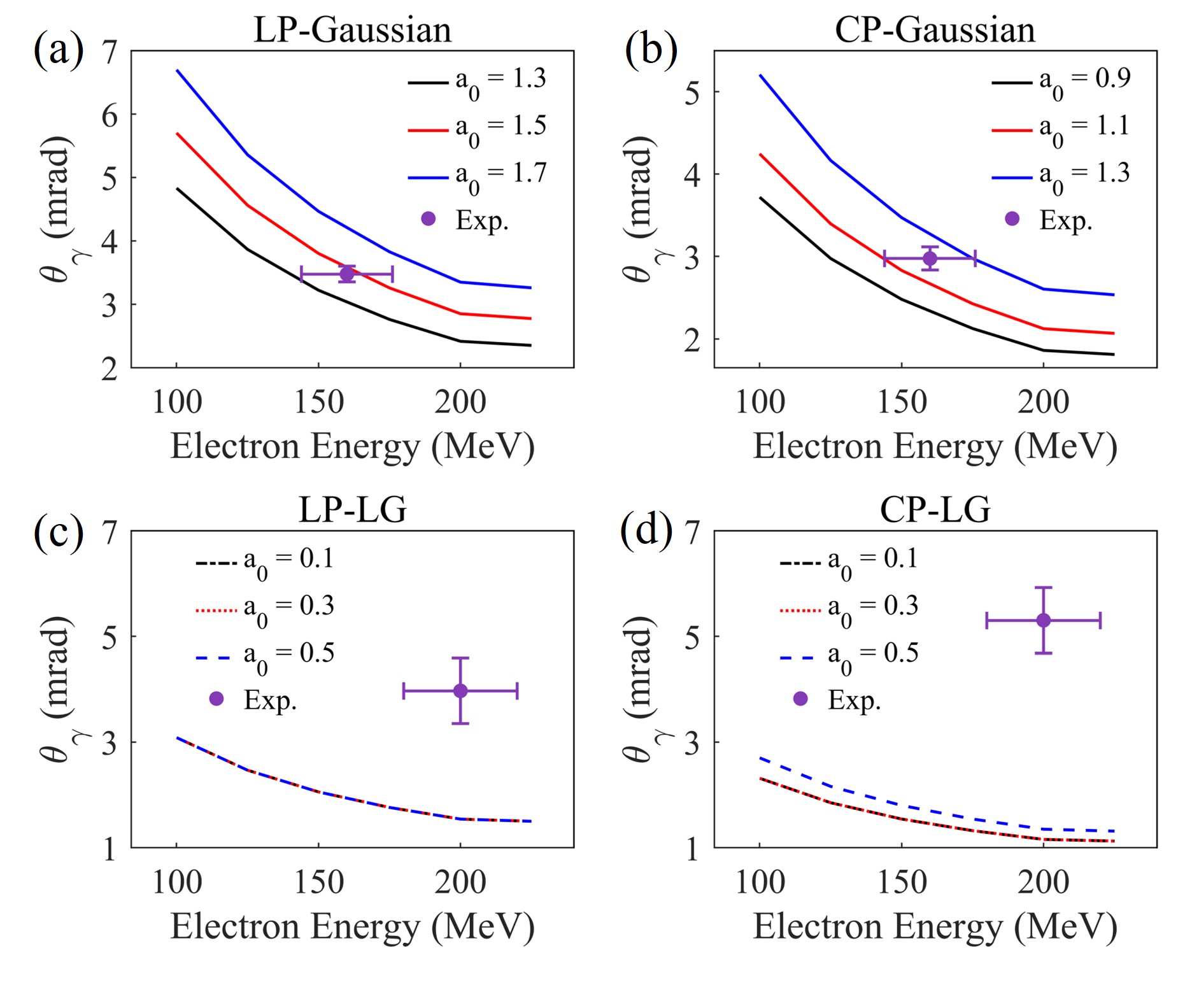}
    \caption{The comparison of divergence angle of $\gamma$ rays between theoretical calculations (details in Appendix C) and experimental results for four colliding lasers: (a) LP-Gaussian, (b) CP-Gaussian, (c) LP-LG, and (d) CP-LG. The solid and dashed lines represent numerical results as a function of $a_0$ and electron energy, while the dots represent the average values of the experimental data [see Fig. \ref{fig3}(e)]. The transverse error bars are due to the electron energy spread, while the longitudinal error bars correspond to the standard deviation of the experimental data.}
    \label{fig4}
\end{figure}

In this Letter, we successfully generate sub-MeV vortex $\gamma$ photons via all-optical inverse Compton scattering of relativistic electrons (hundreds of MeV) colliding with LG laser (intensity $\sim$ \(10^{17} \, \text{W} / \text{cm}^{2}\)). The relativistic electron beam, generated via laser wakefield acceleration \cite{tajimaLaserElectronAccelerator1979,pukhovLaserWakeField2002b,picksleyMatchedGuidingControlled2024c} (LWFA), collides with LG laser of topological charge $l=7$, resulting in the upshift of scattered photon energy to the MeV range. Furthermore, we put forward a novel method to determine the vortex properties by revealing the quantum opening angle of vortex $\gamma$ photons. In quantum electrodynamics theory \cite{ivanovPromisesChallengesHighenergy2022a,ababekriVortexPhotonGeneration2024,liUnambiguousDetectionHighEnergy2024a,luManipulationGiantMultipole2023a,jiangManipulationSpinOrbital2024}, a vortex particle can be viewed as a coherent superposition of plane waves symmetrically distributed on a conical surface defined by the opening angle in momentum space (see Fig. \ref{fig1}). Hence, the ensemble angular divergence of vortex $\gamma$ photons will be larger than those of photons without OAM. In the experiment, we clearly observed the angular divergence of $\gamma$ rays generated by the LG laser exhibits a 1.8-fold anomalous expansion compared to that produced by non-vortex Gaussian laser. This expansive angular divergence is not induced by the divergence angle and Lorentz factor of electron (see Fig. \ref{fig2}).
This phenomenon is interpreted as the overall effect of quantum opening angle of individual vortex $\gamma$ photons (see Fig. \ref{fig3}), lying beyond classical theory without quantum effects, as evidenced by Fig. \ref{fig4}. Our measurement provides the first experimental evidence for the generation of high energy vortex  $\gamma$ photons.

\textit{Experimental setup}. A schematic of the experimental setup\cite{chenPlatformAllopticalThomson2024} is shown in Fig. \ref{fig1}. The amplified high energy laser pulse was split into two components: driving laser and colliding laser. The driving laser, used for LWFA to generate relativistic electrons, delivered 2.6 J of energy with a pulse duration of 25 fs. The driving laser was focused by an F/20 off-axis parabola (OAP) mirror onto the center of a 1 mm × 4 mm supersonic gas jet. The focal spot, with a full-width at half-maximum (FWHM) of 20 $\mu$m, corresponds to a peak intensity of about \(8.3 \times 10^{18} \, \text{W} / \text{cm}^{2}\) and the normalized vector potential \( a_0 \approx 2.0 \). The enclosed energy at the FWHM of the focal spot was approximately 25\%. The LWFA was operated in the ionization injection regime \cite{claytonSelfGuidedLaserWakefield2010} using a gas target composed of 99.8\% helium and 0.2\% nitrogen at a backing pressure of 4 bar, which resulted in a gas density of $\sim$\(2.4 \times 10^{18} \, / \text{cm}^{3}\).  

The focal spots of different colliding lasers are shown in Figs. \ref{fig1}(b) and \ref{fig1}(c). The focal spot size was measured as 8 $\mu$m within FWHM, and with the energy of 260 mJ, the Gaussian colliding laser can reach a peak intensity of about \(5.2 \times 10^{18} \, \text{W} / \text{cm}^{2}\)  (\( a_0 \approx 1.5 \)
 ). For the LG laser, a topological charge of $l=7$ was introduced by replacing the reflective mirror with a 2-inch diameter spiral phase plate supported by a copper substrate \cite{wangHollowPlasmaAcceleration2020}. The focal spot of the LG laser is shown in Fig. \ref{fig1}(c), where the focal spot exhibits a doughnut-shaped distribution with the energy of $\sim$115 mJ. The outer ring diameter at the focal position was approximately 41 $\mu$m, while the inner ring diameter was 27 $\mu$m, corresponding to a peak intensity of about \(1.5 \times 10^{17} \, \text{W} / \text{cm}^{2}\). The slight asymmetry in the intensity distribution of focal spot arises from the residual intensity and phase nonuniformities in the laser \cite{denoeudInteractionUltraintenseLaser2017,wangHollowPlasmaAcceleration2020}, as well as potential engineering imperfections in the phase mask and the broad spectrum of the laser. A fork-shaped pattern (details in Appendix A) formed by the interference between the Gaussian driving laser and the LG colliding laser showed the vortex phase of the LG laser, which is consistent with the experimental design \cite{suedaLaguerreGaussianBeamGenerated2004a}.
 
\textit{Results of low-divergence electron beams}. The energy and divergence of the electron beam are measured using an electron spectrometer.
The spectrometer consisted of a 0.98-T dipole magnet for dispersion and a phosphor screen, viewed by a 16-bit CCD camera. Two different types of electron spectra are applied during the Gaussian and LG colliding lasers, as shown in Figs. \ref{fig2}(a) and \ref{fig2}(b), respectively. The relativistic electrons interacting with LG lasers possess higher central energy compared to Gaussian laser. The averaged quasi-monoenergetic peaks of these two types of electrons are located at 160 MeV for Gaussian and 200 MeV for LG lasers, and both of these have an energy spread of approximately 15$\sim$25\%. Fig. \ref{fig2}(c) displays the probability plot of the divergence angles of the relativistic electrons in the transverse direction after passing through the electron magnetic spectrometer. It shows that the average divergence angles of the electrons are 2.0 mrad and 1.2 mrad for the Gaussian and LG lasers, respectively. These differences of the electrons stem from the LWFA with different drive beams, as we selected two different locations for the pickup of the colliding laser, making the leftover driving laser slightly different (details in Appendix A), and thus the electrons for these two cases also differed. It should be noted that these differences of the electrons existed prior to the collision.

Figure \ref{fig2}(d) shows the quasi-monoenergetic $\gamma$ rays energy spectra measured by Range Filter \cite{hojbotaHighenergyBetatronSource2023}, where both colliding lasers are in LP mode. The scattered photons are boosted to $E_\gamma \approx 2 \gamma_e^2(1-\cos \theta) \hbar \omega_L$, where $\theta$ is the interaction angle, $\gamma_e$ and $\omega_L$ are the Lorentz factor of electron and laser central frequency, respectively. The resulting $\gamma$ rays exhibit a quasi-monoenergetic spectrum, with a central peak energy of 0.8 MeV. When the colliding laser is replaced with a Gaussian laser, the peak energy of $\gamma$ rays decreases to 0.5 MeV, which is attributed to a reduction in the central energy of the electron beam, as shown in Figs. \ref{fig2}(a) and \ref{fig2}(b). 

\textit{Observation of $\gamma$ rays angular distribution}. Typical $\gamma$ rays angular distributions are shown in Figs. \ref{fig3}. It can be seen from Figs. \ref{fig3}(a) and \ref{fig3}(b) that the angular distribution patterns of the $\gamma$ rays exhibit a faintly elliptical shape for the LP-Gaussian and a circular profile for the CP-Gaussian colliding laser (details diagnosis in Appendix B). For equivalent laser energy, LP beams generate transverse electric fields enhanced by $\sqrt2$ relative to CP beams. This field enhancement drives nonlinear scattering dynamics in LP-Gaussian interactions, manifesting as polarization-direction elongation of $\gamma$ rays angular distributions [Fig. \ref{fig3}(a)].  This elongation was attributed to the increased amplitude of the figure-eight trajectory oscillations of the electron within its average rest frame \cite{yanHighorderMultiphotonThomson2017,milburnElectronScatteringIntense1963,sarachikClassicalTheoryScattering1970}.

When the colliding laser is an LG laser, the angular distribution did not show the expected doughnut-like shape with reduced intermediate intensity. This discrepancy may arise from two key factors: the use of femtosecond pulses in our experiment, in contrast to the picosecond long pulse considered in Ref. \cite{petrilloComptonScatteredXGamma2016}, and the non-purity vortex state of the topological charge in the LG laser employed, which may influence the OAM distribution of the $\gamma$ photons.

Even so, by quantifying the angular distribution for four colliding lasers, we observed an interesting phenomenon. That is, the angular distribution of $\gamma$ rays produced by LG laser exhibits an anomalous expansion compared to Gaussian laser as shown in Fig. \ref{fig3}(e). The divergence angle of $\gamma$ rays is defined as $\theta_\gamma=\sqrt{\theta_m{ }^2-\theta_e{ }^2}$, where $\theta_m$ is the measured value of angular distribution of $\gamma$ rays , which is defined as half of the maximum divergence angle at $1/e$ intensity. Employing this formula, we quantify and eliminate the impact of electron divergence angle $\theta_e$ on the angular distribution of $\gamma$ rays. The mean value of $\theta_\gamma$ for all shots is 3.5 ± 0.2, 3.0 ± 0.2, 4.2 ± 0.6, and 5.3 ± 0.7 for LP-Gaussian, CP-Gaussian, LP-LG, and CP-LG lasers, respectively. 
Therefore, when the colliding lasers are LG lasers, an interesting anomalous expansion of divergence angle of the $\gamma$ rays is observed, which is the result of quantum opening angle of the vortex photons and cannot be predicted by classical understanding. Compared to LP-LG, CP-LG laser preserves rotational symmetry, facilitating the transition of OAM and consequently exhibiting a larger angular distribution \cite{ivanovPromisesChallengesHighenergy2022a,ababekriVortexPhotonGeneration2024}.

\textit{Vorticity analysis of $\gamma$ photons}. The reasonable inference that scattered photons carry OAM stems from three experimental facts. Firstly, as shown in Fig. \ref{fig2}, compare to Gaussian laser, the colliding electrons in LG laser possess a higher $\gamma_e$. According to the theory of single electron radiation \cite{sarachikClassicalTheoryScattering1970,cordeFemtosecondRaysLaserplasma2013}, if the emitted $\gamma$ photons do not carry OAM, the overall angular distribution of the $\gamma$ rays in LG mode should be confined within a cone of aperture angle approximately \(1/\gamma_e\), which is narrower than that generated by Gaussian laser. Secondly, when the colliding laser is LG mode, the electron bunch exhibits a smaller average divergence angle of 1.2 ± 0.3 mrad, compared to 2.0 ± 0.3 mrad for the Gaussian mode, as illustrated in Fig. \ref{fig2}(c). The contribution of the electron beam divergence angle would result in a reduced angular distribution of the $\gamma$ rays in LG mode. Thirdly, in the experiment, the $a_0$ is significantly lower for the LG laser, and the nonlinear effect is much weaker compared to Gaussian mode, without leading to an expansion in $\theta_\gamma$ \cite{yanHighorderMultiphotonThomson2017}. Therefore, the anomalous expansive angular distribution lies beyond usual understandings. In the experimental configuration with $l = 7$, the scattered photons carry OAM due to the conservation of angular momentum. As a whole, the angular distribution of \(\gamma\) rays exhibit an expansion compared to Gaussian mode, as shown in the inset of Fig. \ref{fig1}(a).

To further confirm the vortical properties of $\gamma$ photons, we conducted theoretical calculations using two different codes \cite{cordeFemtosecondRaysLaserplasma2013,jacksonClassicalElectrodynamicsThird1998} for vortex laser ($l = 7$) and non-vortex laser (recovers the Gaussian laser by setting $l=0$). The radiation emitted by individual moving charged particle can be calculated using the Liénard-Wiechert potential (details in Appendix C). 
Figure \ref{fig4} shows a comparison between theoretical calculations and experimental results for four colliding lasers as functions of the electron energy $\varepsilon$ and the $a_0$. 
As visualized in Fig. \ref{fig4}(a), for the LP-Gaussian laser with $a_0=1.5$, the experimental results agree well with theoretical predictions. For the CP-Gaussian laser in Fig. \ref{fig4}(b), after accounting for the electron beam energy spread and fluctuations in the $a_0$, the experimental results also agree with the theoretical calculations. The fluctuations in $a_0$ arise from the angular misalignment of the quarter-wave plate and the efficiency of polarization conversion. 
Within the experimental parameter range (where \(a_0 = 0.3\) and $\varepsilon = 200 \pm 20$ MeV) for the LP-LG, the $\theta_\gamma$ was approximately \(4.2\) mrad. This value is larger than the \(1.6\) mrad predicted by theoretical calculations and lies beyond the experimental error margin, as clearly demonstrated in Fig. \ref{fig4}(c). 
For CP-LG, the experimental results still deviate from the calculations as exemplified in Fig. \ref{fig4}(d). 
In brief, the theoretical calculations for non-vortex Gaussian lasers align well with the experimental observations, whereas for vortex lasers, the experimental results are significantly larger than the classical theoretical predictions. This discrepancy is particularly significant because numerical simulations based on the electric field derived from the paraxial approximation \cite{loudonTheoryForcesExerted2003,davisTheoryElectromagneticBeams1979} fail to reproduce the experimental results of LG laser. The main reason for this discrepancy is that calculations based on electron trajectories do not account for the opening angle of the momentum cone associated with vortex photons \cite{ivanovPromisesChallengesHighenergy2022a,ababekriGenerationUltrarelativisticVortex2024a}.

Please note that the direct measurement of OAM remains fundamentally challenged by the incoherent nature of $\gamma$ photons generated via inverse Compton scattering, although there is a proposal to probe the OAM of $\gamma$ photons via the isovector giant quadrupole resonance \cite{luManipulationGiantMultipole2023a}. 
Moreover, quantum calculations are also difficult to apply to such $\gamma$ rays due to the complexity of these multiparticle interactions.
Building on the previous analysis, our method provides experimental evidence for the vorticity by measuring the expansive angular distribution of $\gamma$ rays induced by the quantum opening angle.

In conclusion, we have successfully generated sub-MeV vortex $\gamma$ photons for the first time, and established a novel approach for determining high energy vortex $\gamma$ photons by revealing the quantum opening angle. Our result provides new experimental evidence for the generation of vortex $\gamma$ photons and offers a new perspective for exploring previously inaccessible quantum mechanical properties in OAM-dependent phenomena. This breakthrough paves the way for studying OAM-induced quantum phenomena \cite{ababekriVortexPhotonGeneration2024,zilgesPhotonuclearReactionsBasic2022a}, having significant impact for nuclear physics, nuclear astrophysics, strong field physics, etc.

\vspace{1em}
\begin{acknowledgments}
\textit{Acknowledgments—}We thank Donald Umstadter from University of Nebraska, Lincoln, Vittoria Petrillo from University of Milan, Liangliang Ji and Shiyu Liu from Shanghai Institute of Optics and Fine Mechanics, Yaojun Li, Wenzhao Wang 
from Shanghai Jiao Tong University for useful discussion. This work was supported by the National Key Research and Development Program of China (Grant
No.2021YFA1601700, No. 2024YFA1610900), the National Natural Science Foundation of China (Grant No. 12074251, 11991073, No. 12425510, No. U2267204, No. 12441506, No. 12225505), the Strategic Priority Research Program of the Chinese Academy of Sciences (Grants No. XDA25010500, No. XDA25030400, No. XDA25010100). 
The computations in this paper were run on the $\pi$ 2.0 and Siyuan cluster supported by the Center for High Performance Computing at Shanghai Jiao Tong University. The authors would like to acknowledge the Yangyang Development Fund, China.
\end{acknowledgments}


\bibliographystyle{apsrev4-2}
\bibliography{mycit}

\begin{figure}[H]
    \centering
    \includegraphics[width=1\linewidth]{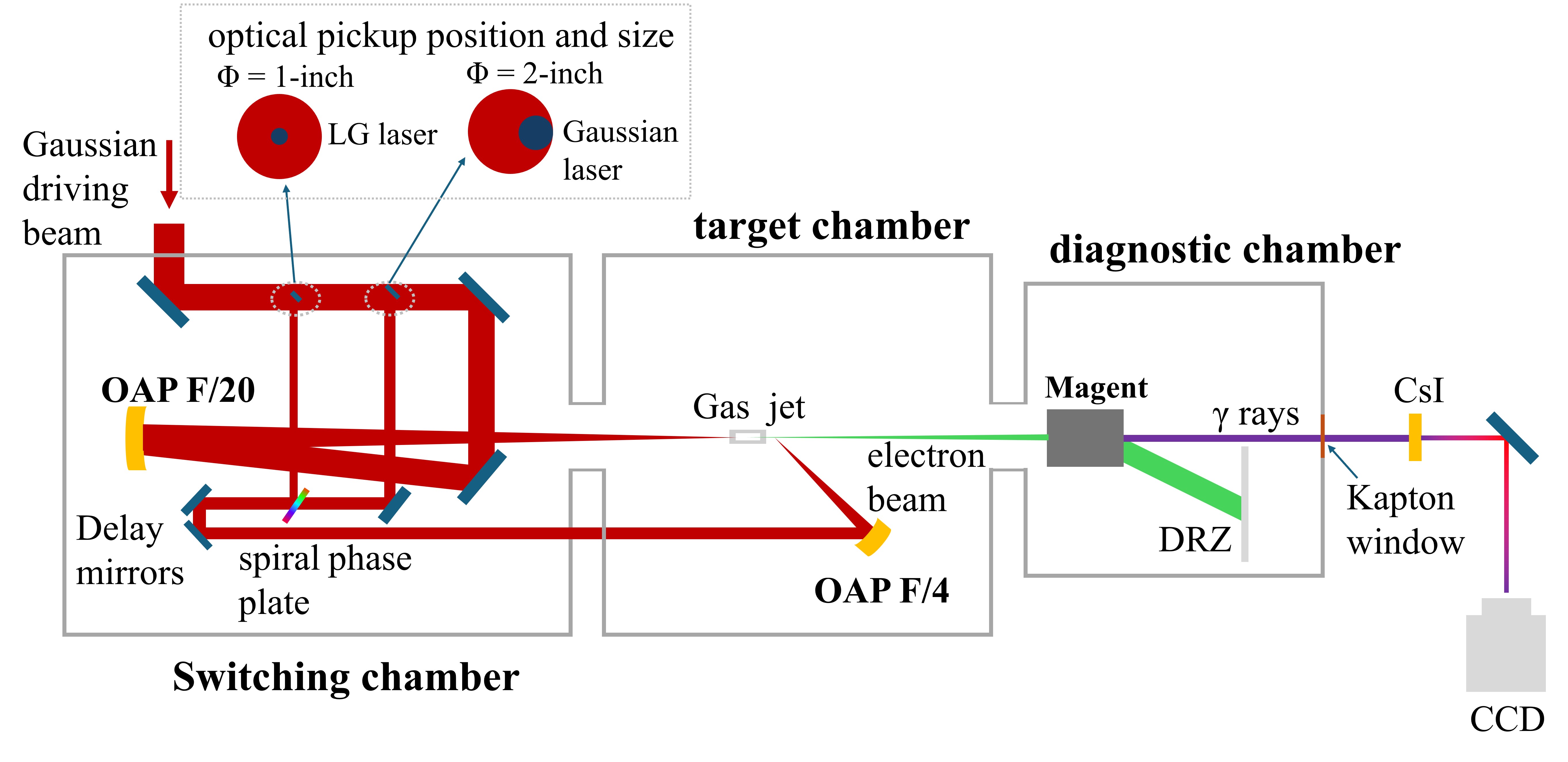}
    \caption{Difference picking position for LG colliding laser and Gaussian laser.}
    \label{s1}
\end{figure}

\vspace{1em}
\appendix
\textit{Appendix A: Scattering laser beams}. In our inverse Compton scattering experiment, all colliding beams were picked up from the main driver laser pulse after amplification, using an elliptical mirror of varying sizes. The colliding laser laser beam was focused using an F/4 OAP and propagated within the plane of the electron beam, with a 135° collision angle relative to the propagation axis of the relativistic electron beam. In the case of the Gaussian laser, the picking position was located at the edge of the main driving laser beam, which has a diameter of 2 inches and contains approximately 260 mJ of energy as shown in Fig. \ref{s1}. For the LG mode, the picking position was optimized at the beam center to ensure the highest vortex quality, with the diameter reduced to 1 inch and approximately 115 mJ of energy to align the intensity peak with the phase singularity. So, for the case of Gaussian colliding lasers, the inverse Compton scattering process occurs within the nonlinear regime. In contrast to the Gaussian colliding laser, the LG colliding laser is extracted from the center of the driving laser, which causes the remaining laser to excite the transversely symmetric wakefield inside the plasma, thus helping to reduce the divergence angle of the electrons generated by laser wakefield acceleration. 

\begin{figure}[H]
    \centering
    \includegraphics[width=0.8\linewidth]{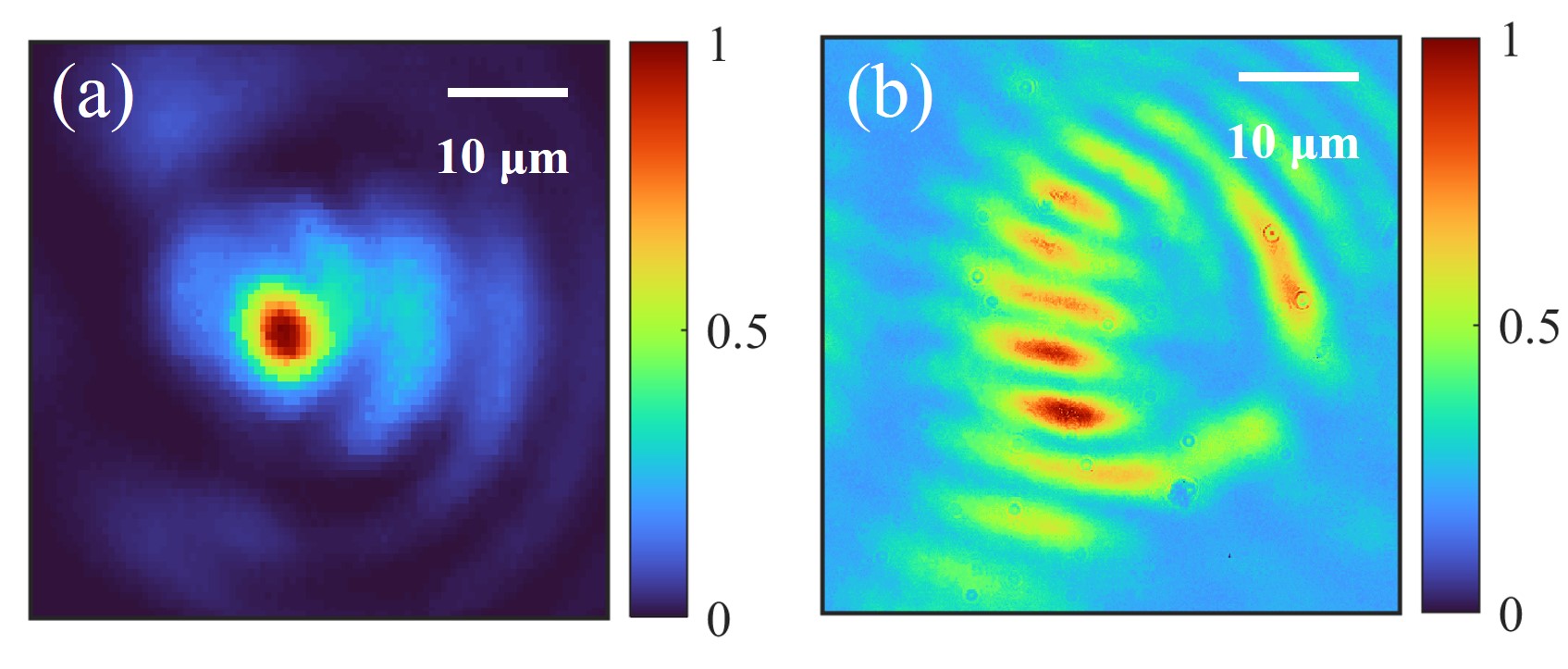}
    \caption{The interference patterns of the driving laser and different colliding laser: (a) Gaussian and (b) LG, respectively.}
    \label{s2}
\end{figure}

An imaging system was used to record the colliding laser beam profile at the focal spot. Figs. \ref{s2}(a) and \ref{s2}(b) show the interference patterns of the Gaussian colliding laser and the LG colliding laser interfere with another same Gaussian beam, respectively. While the Gaussian mode showed the Newton rings in the interference pattern, a fork-shaped interference pattern was shown in LG mode, which provided the evidence of orbital angular momentum. The number of the extra fringes are 8, which indicate that the LG colliding laser has topological charge of $l=7$, which is consistent with the phase plate design.

\begin{figure}[H]
    \centering
    \includegraphics[width=0.7\linewidth]{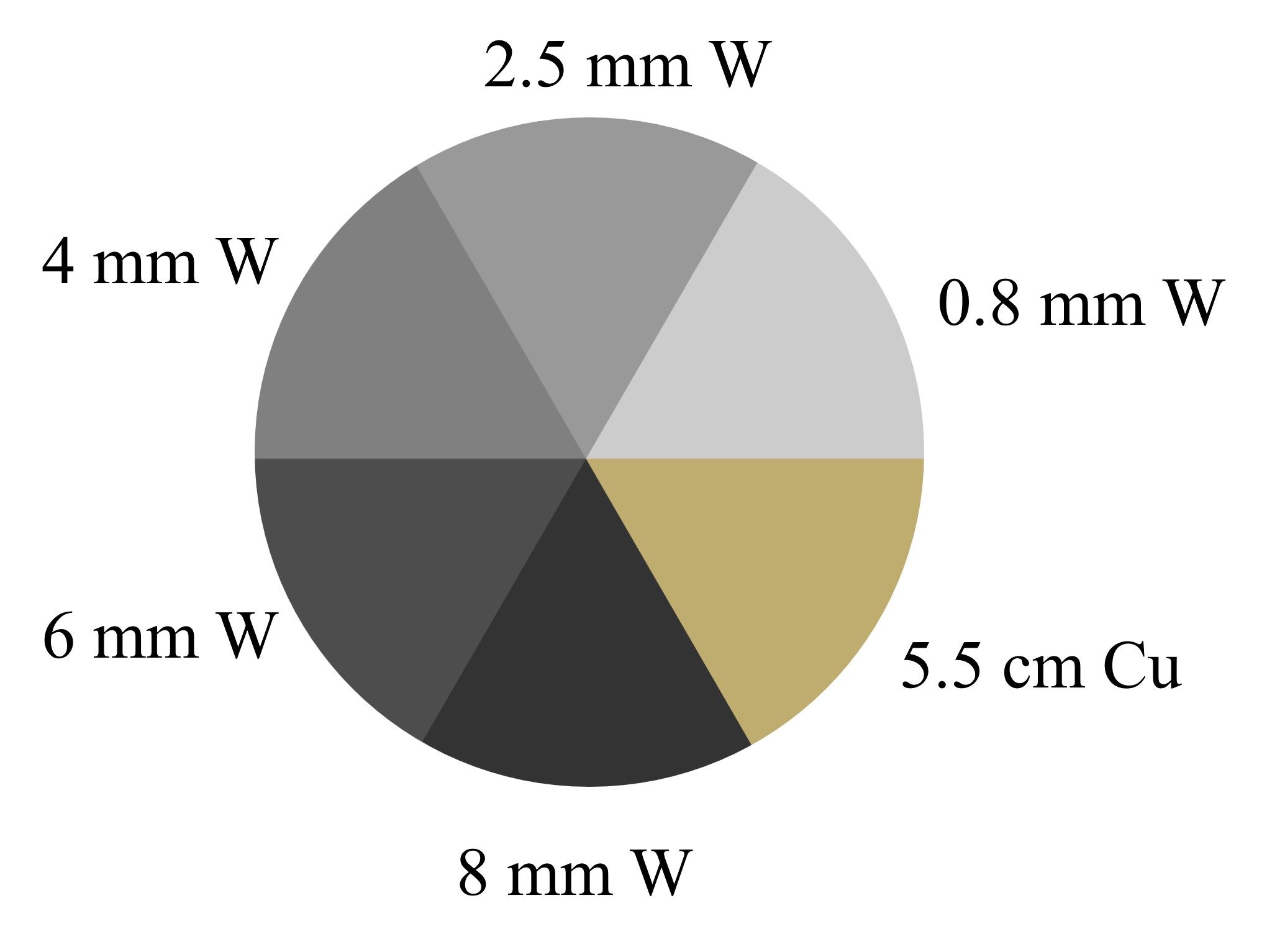}
    \caption{The schematic diagram of the Range Filter stack.}
    \label{s3}
\end{figure}

\textit{Appendix B: $\gamma$ rays diagnosticss}. In the experiment, $\gamma$ rays energy spectra within the range of hundreds of keV were measured using a Range Filter. Tungsten and copper materials of varying thickness were used for this measurement, as depicted in Fig. \ref{s3}. To reconstruct the $\gamma$ rays spectrum from the experimental data, a gradient-descent-based iterative optimization algorithm was applied. During the iterative process, a normalized squared error function was adopted as the error criterion. Through continuous iterations, the difference between the actual and predicted signals was minimized until the error converged to a minimum value. After the iterations were completed, the resulting parameters were smoothed and scaled to consider the total photons count. This process yielded the final incident $\gamma$ rays spectrum.

The spatial characteristics of the $\gamma$ rays were measured using a calibrated CsI array scintillator, consisting of an 80×80 array of 1×1×10 mm voxels, imaged in real-time using a 14-bit electron-multiplying CCD camera. The CsI array was placed outside the vacuum chamber, aligned along the axis of the electron beam, and positioned 2.0 m from the interaction point. To shield the detector from spurious radiation and improve the signal-to-noise ratio, the array was encased in side-shielded area by lead. In our experimental setup, betatron radiation co-occurring with electrons during laser wakefield acceleration could be generated. Using a 10 mm thick CsI scintillator ensures that the signals observed originate from the inverse Compton scattering process.

\textit{Appendix C: Theoretical calculation}. 
Within the framework of classical electrodynamics, we employed two theoretical approaches to interpret the angular distribution of the radiation photons in the experiments. One was based on a self-compiled code utilizing the Liénard-Wiechert potential \cite{cordeFemtosecondRaysLaserplasma2013,jacksonClassicalElectrodynamicsThird1998}, and the other was the Virtual Detector for Synchrotron Radiation (VDSR) \cite{chenModelingClassicalQuantum2013}. In the calculation, the LG laser field is written as \cite{wangHollowPlasmaAcceleration2020,zhouGammarayVortexBurst2023}:
\[
\vec{a}(\rho, \phi, x, t)=u_p^{|l|}(\rho, x) e^{i \Theta_{k l p}(\rho, \phi, x)} e^{-i \omega t}(\hat{y}+\sigma i \hat{z}),
\]
where \(\sigma = \pm1\) represents circular polarization, while \(\sigma=0\) denotes linear polarization. The amplitude term is given by:
\[
u_p^{|l|}(\rho, x)=a_0 \frac{C_{|l| p}}{\left(1+x^2 / x_R^2\right)^{1 / 2}}\left(\frac{\sqrt{2} \rho}{w(x)}\right)^{|l|} L_p^{\left | l\right |} e^{-\frac{\rho^2}{w^2(x)}},
\] 
and the phase term by:
\[
\Theta_{k l p}(\rho, \phi, x)=k x+l \phi-(2 p+|l|+1) \tan ^{-1} \frac{x}{x_R}+\frac{k \rho^2 x}{2\left(x^2+x_R^2\right)},
\]
where $w(x)$ is the beam radius at position $x$, $a_0$  is the normalized laser vector potential amplitude, $L_p^{|l|}$  are the generalized Laguerre polynomials, and  $l$ is the topological charge, while  $p$ and $x_R$  represent the radial index and the Rayleigh length, respectively.

\begin{figure}
    \centering
    \includegraphics[width=1\linewidth]{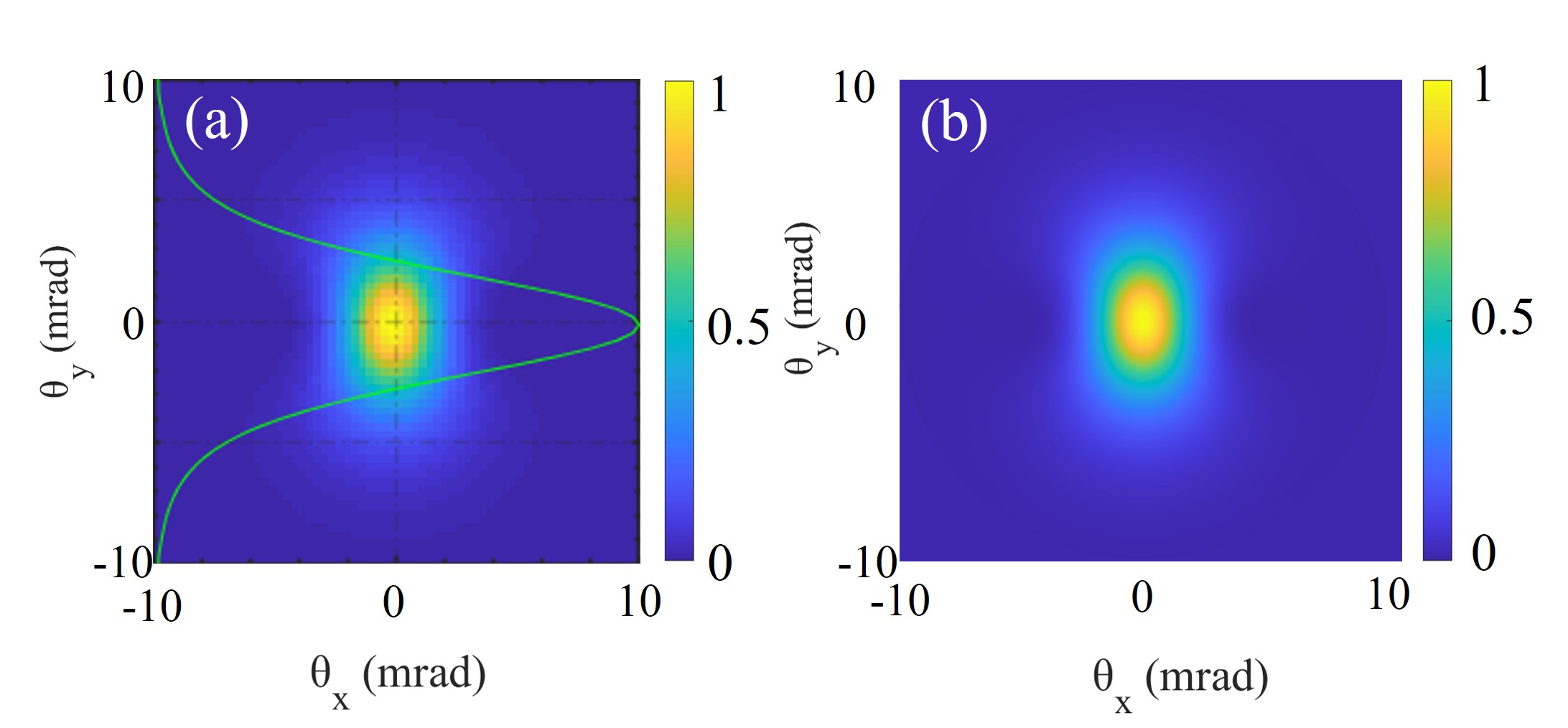}
    \caption{The angular distribution of the radiation photons calculated using two different codes for LP-LG colliding laser mode ($a_0 = 0.3$ and $\varepsilon = 200 \pm 20$ MeV): (a) Self-compiled code based on Ref. \cite{cordeFemtosecondRaysLaserplasma2013} and (b) VDSR code.}
    \label{s4}
\end{figure}

The analytical calculations, shown in Fig. \ref{s4}, were carried out by considering the angle of incidence, the laser temporal profile, and the laser $a_0$ distribution (or intensity distribution) from the experiment. In the framework of classical electrodynamics, the radiation emitted can be calculated using the Liénard-Wiechert potential \cite{cordeFemtosecondRaysLaserplasma2013}:
\[
\mathbf{E}_{\mathrm{rad}}(\mathbf{r}, t)=\left[\frac{e}{c} \frac{\hat{\mathbf{n}} \times(\hat{\mathbf{n}}-\boldsymbol{\beta}) \times \dot{\boldsymbol{\beta}}}{(1-\boldsymbol{\beta} \cdot \hat{\mathbf{n}})^3 R}\right],
\]
where the $\hat{\mathbf{n}}$, $R$, $\beta$ and $\dot{\boldsymbol{\beta}}$ present the observation direction, observation distance, normalized velocity, and acceleration of the charged particle, respectively. By substituting the electric field expression with the LG form that satisfies Maxwell's equations, the radiation emitted by a single electron in the LG laser field can be calculated. Figure \ref{s4} illustrates the angular distribution of radiation photons calculated using two independent codes. The results demonstrate good agreement between the two codes under varying colliding beam configurations. While the calculations for the Gaussian beam case align closely with the experimental data, those for the Laguerre-Gaussian beam case exhibit significant discrepancies. These results suggest that the interaction of relativistic electrons with LG laser involves unique quantum effects of vortex photons and requires further study.

\end{document}